\documentstyle[12pt,aasms4]{article}
\lefthead{Lambert and Sheffer}
\righthead{Spectroscopy of Epsilon Aurigae}
\begin{document}

\title{Inter-Eclipse Spectroscopic Snapshot of Epsilon Aurigae with {\it HST}}
\author{Yaron Sheffer and David L. Lambert}
\affil{Department of Astronomy, University of Texas, Austin, TX 78712\\
Electronic mail: yaron {\it OR} dll @astro.as.utexas.edu\\
astro-ph/9905129}

\begin{abstract}
A single-epoch low resolution GHRS spectrum of the eclipsing binary
Epsilon Aurigae was obtained while the secondary was orbiting towards eclipse
by the primary. The spectrum as recorded between 1175---1461 \AA\ is rich with
emission and absorption lines which include stellar
and interstellar components. The emission line profiles have the appearance of
double-peaked emission with a stronger red component at a radial velocity
of +108 km~s$^{-1}$,
an overlying unresolved absorption component at $-$20 km~s$^{-1}$
and a weaker blue emission bump at ca. $-$92 km~s$^{-1}$.
We compare these observational results with known orbital properties of
the $\epsilon$~Aur binary system, and propose that the emission originates
at the inner radius of the disk surrounding the enigmatic secondary.
We interpret the kinematic data as a possible means to uncover the underlying
stellar masses and we speculate about the binary's relationship to other
``high-mass'' models. 
\end{abstract}
\keywords{binaries: eclipsing --- line: profiles --- stars: individual
($\epsilon$~Aur) --- techniques: radial velocities --- ultraviolet emission}

\section{INTRODUCTION}
The EA-type variable star Epsilon Aurigae (HD 31964, HIC/P 23416, spectral
type F0 Ia) exhibits fadings of about
0.8 mag in $V$ every 27.1 years, which are attributable to eclipses by an
invisible flattened disk of ``dark matter'' (Huang 1965, Wilson 1971).
The nature of the disk and the identity of whatever is lurking at its center
are so far undetermined in spite of observations extending back
to the 1824 work of Fritsch (as related by Wood 1985).
With only 4 eclipses occurring per century, each 2-year long eclipse
plays to an observational crowd of a different generation, at times equipped
with a dramatically improved instrumentation.
The last one of 1982--4 was the first to be monitored by spacecraft
in the ultraviolet (UV), predominantly via low-resolution spectroscopy done
with the {\it International Ultraviolet Explorer (IUE)\/}.

Indeed, during the last eclipse enough observational data were obtained
to enable some crude modeling of a giant disk of gas and dust in orbit around
the F type supergiant (see, e.g., Kemp et al. 1985, Ferluga and Hack 1985,
Backman 1985, Lambert and Sawyer 1986, Sait\={o} et al. 1987). The data
included shell absorption lines, excess ultraviolet continuum flux, black-body
fitting of infrared fluxes, and polarization measurements. The current consensus
about the secondary star promotes a doughnut-shaped disk with a central
clearing around the star, see e.g., Carroll et al. (1991) and Lissauer et al.
(1996), in order to account
for the remarkable mid-eclipse brightening of $\epsilon$~Aur. It is not clear
how the hole is maintained since no observational data so far has ever been
directly linked to any object that may reside at its center (but see Lissauer
and Backman 1984 and Eggleton and Pringle 1985 for proposing $\epsilon$~Aur
as a triple stellar system).

Since the system is
visible throughout its 27-year long cycle, it deserves scrutiny at all times,
especially in the UV where the secondary may be detectable (Hack and Selvelli
1979, Parthasarathy and Lambert 1983).
For this reason we have observed $\epsilon$~Aur with the GHRS on the {\it Hubble
Space Telescope (HST)\/} at a time intermediate between two primary
eclipses (the next one
will be centered on the year 2010). Our principal goal was to detect evidence
of the enigmatic secondary. The now-retired GHRS provides data of a quality far
surpassing that achieved with the {\it IUE\/}
satellite. In this paper we describe  and analyze our GHRS spectrum
and interpret our data by comparison
with the predicted orbital configuration of the binary.

\section{OBSERVATIONS AND REDUCTIONS}
The {\it HST\/} was trained onto $\epsilon$~Aur under program
GTO 6289 (with DLL as PI) as part of Cycle 5. Using the G140L grating to
capture wavelengths between 1174.8---1461.2 \AA, 13 exposures of equal exposure
time and totaling 3536-sec were acquired starting at 09:26:54 UT on
1996 February 16. These were taken through the Small Science Aperture
(SSA, 0.22\arcsec\ $\times$ 0.22\arcsec\ in size) to provide a
nominal resolving power of $R = \lambda/\Delta\lambda$ = 2000 (at 1300 \AA).
Normal reduction routines have been followed
by processing the spectrum in the {\sc stsdas} environment of {\sc iraf}, where
we combined the 13 sub-exposures into a single spectrum. Each sub-exposure
was acquired four times but with the spectrum shifted by 1/4-diode steps
yielding about 3.7 pixels per resolution element.

Flux and wavelength calibrations were performed by the
Space Telescope Science Institute (STScI) pipeline. However, for precise
radial velocity work, refinements of the zero offset
were secured with the help of an acquired Pt/Ne lamp spectrum
on-board the {\it HST\/}. Indeed, on closer inspection,
the calibration lines were found to be shifted away from their expected rest
wavelengths. The task of measuring Pt/Ne line profiles
was complicated by the high fraction of close blends among the observed lines,
as can be seen in the listing of Pt/Ne lines by Reader et al. (1990).
Our selection of 31 isolated or de-blended lines (fitted by Gaussian profiles
in the {\sc iraf splot} environment) yields the following results.
There is a shift to longer wavelengths, $\Delta\lambda$ = +0.117$\pm$0.012 \AA,
and the average full-width at half-maximum (FWHM) of the lines is
0.655$\pm$0.039 \AA, which corresponds to a spectral resolving power of $R =
1980\pm$120 at 1300\AA. No significant dependence
on the wavelength is detected for these parameters from the line sample which
spans the entire spectrum.
In order to derive the correct radial velocity information from the stellar
spectrum we had to correct its STScI-given wavelength scale by subtracting
the wavelength shift present in the Pt/Ne calibration spectrum. Since the
correction due to geocentric motion has already been applied during
the calibration at STScI, the velocities are reported in heliocentric frame.

Due to its distance of about 0.6 kpc on a line of sight only 1\fdg2 from the
Galactic plane, $\epsilon$~Aur suffers appreciable interstellar extinction
in the UV.
We have chosen to de-redden the GHRS spectrum using Seaton's (1979) recipe
and the accepted reddening towards $\epsilon$~Aur, $E(B-V)$ = 0.30 (Hack and
Selvelli 1979). After
correction, the flux is higher by 3.2 and 2.5 mag at 1174 and 1461 \AA,
respectively, i.e., a differential enhancement of the spectrum's blue end
by $\times$1.9 relative to its red end. All our flux measurements reported below
are from the extinction-corrected data. No correction is made for additional
reddening from dust in the binary system that might be affecting
the observed spectrum.

\section{EMISSION LINE INVENTORY}
The fully reduced, co-added, and  dereddened spectrum is shown in Figure 1.
It exhibits a sloping continuum level that is higher
at redder wavelengths, upon which about a couple of dozen emission lines
are superimposed.
The emission lines are readily discernible below about 1350\AA.
We have identified 18 transitions that appear in emission in the spectrum
of $\epsilon$~Aur. The identified lines are all transitions to ground and
very low-excitation ($\leq$ 0.04 eV)
states of abundant atoms and ions: H I, C II, N I, O I, Si II,
and S II. In the main, species not listed but which might be expected
do not have strong resonance lines in the observed region.
At the reddest one-third of the spectrum it is difficult to tell
if emission lines are present due to the emission-like appearance of continuum
gaps between absorption lines of the F0 supergiant.

\subsection{Profiles and Velocities}
Table 1 lists for each line its species identity,
multiplet number of said species,
the rest wavelength, the shift-corrected observed wavelength,
the derived radial velocity, value of $A_{ij}$, observed flux and
its error bar
as determined by summing all pixels within the line profile and above
the continuum in both the spectrum itself and in the flux uncertainty
distribution (shown as the lower curve in Fig. 1), and finally, the FWHM
of the emission lines. Note that the observed wavelengths
have been corrected by $-$0.117 \AA, as described in \S2.
Determinations of line
centroids for the observed wavelengths and of line FWHM values have been
accomplished by fitting of Gaussian profiles within {\sc splot}.
All atomic transition parameters were taken from Smith et al. (1996).

On inspection, it is apparent that the emission lines' profile comprises
a strong red peak and a weak blue peak separated by a central absorption.
The three panels in Figure
2 demonstrate the shape of the emission line profiles for (a) Si II lines
around 1190 \AA, (b) S II lines around 1250 \AA, and (c) the three lines
of O I, Si II, and C II at 1302, 1309, and 1334 \AA, respectively.
For proper graphic comparison we have subtracted the local continuum
for each line and then re-scaled the peak flux to unity.
In general, towards the red there is a sharp
drop that reaches the continuum at a radial velocity of about +275$\pm$25
km~s$^{-1}$,
while the more gradual decline to the blue terminates at about
$-$200$\pm$25 km~s$^{-1}$. Therefore, the full-width at zero-intensity
(FWZI) is about
475$\pm$35 km~s$^{-1}$ with an average radial velocity at about +37$\pm$18
km~s$^{-1}$.

In Table 2 we list all emission transitions that provided a clean and
clear view of the blue peak and/or the central absorption, together with
radial velocities and component widths that were extracted by means of
multi-component Gaussian profile fits within {\sc splot}.
The central absorption features exhibit different strengths
(more about this absorption in \S4.1).
The blue peaks are weaker than the red peaks: on average, they are 0.27$\pm$0.14
of the red-side flux. Such a state of affairs makes these blue peaks more
susceptible to noise and they are hard to distinguish when transitions
are starting to blend. Hence, the number of
emission profiles (in Table 2) which are suitable for blue peak and central
absorption analysis is only about half of the number that is available for
red peak analysis from Table 1.

The average radial velocity from 16 red emission peaks
(from Table 1 but without H I) is +108$\pm$18 km~s$^{-1}$.
We also did not include the 1260\AA\ Si II line due to blending
with absorption from C I on its blue side. The emission lines
appear to have a FWHM which is narrower than that of the Pt/Ne lamp lines:
the mean FWHM of 15 stellar lines is 0.63$\pm$0.10 \AA, not including the
known blends of $\lambda\lambda$ 1264 and 1304. This is in good agreement
with {\it HST\/} performance where the stellar profile is ca. 95\%
the width of the SSA FWHM (Stevens 1998).
From Table 2 data, the following averages emerge. The radial velocity for
nine blue peaks is $-92\pm$21 km s$^{-1}$ and their average FWHM is
0.56$\pm$0.09 \AA. Values for the central absorption features are
$-20\pm$14 km s$^{-1}$ and 0.50$\pm$0.15, respectively.

The only emission line not attributable to $\epsilon$~Aur is Ly$\alpha$ of H I.
It is an expected contamination from the Geo-coronal background observed
through the GHRS aperture.
Luckily, the SSA is significantly smaller than the {\it IUE\/}
apertures so that the Ly$\alpha$ emission seen here is not strong enough to
mask wide parts of the spectrum, as happened with {\it IUE\/} spectra.
This emission line is the only spectral feature to vary appreciably in strength
among the GHRS sub-exposures; its derived ``error'' computed by {\sc stsdas}
is an indicator of the true variability and is
a solid proof of Geo-coronal origin. As such, this line also serves as an
indispensable check on our radial velocity scale. The line is within 0.1 pixels
(or +4 km~s$^{-1}$) of the predicted position.
The Geo-coronal H I emission line plays another supporting role by
exhibiting a single Gaussian component, thus showing that
the other emission lines with more complex profiles
are definitely signatures of velocity fields within the $\epsilon$~Aur
system and of absorbing matter far from Earth.
Here, Ly$\alpha$ emission is found inside a wide, saturated absorption feature,
of interstellar origin.

\subsection{The O I Lines from Multiplet 2}
The complex of ``four'' emission lines between 1302\AA\ and 1309\AA\
(see Figure 1 and Table 1) is composed of three O I lines of multiplet 2 and
two Si II lines of multiplet 3. Only four lines out of five are readily
visible due to blending of O I 1304.86\AA\ with Si II 1304.37\AA.
In the low-resolution {\it IUE\/} spectra, the entire group
of transitions were seen as a single unresolved feature. Furthermore,
we should keep in mind that during the 1983 eclipse there was
no weakening of the $\lambda$1305 complex (Parthasarathy and
Lambert 1983). Together with the UV continuum below $\sim$1400\AA, the
two cannot be associated with the eclipsed portion of the primary's face, but
rather they originate in the vicinity of the non-eclipsed secondary or in a
chromosphere-like region larger than and around the eclipsed hemisphere
the F0 supergiant, as described by Parthasarathy and Lambert.

An interesting story is told by the relative fluxes of the O I lines.
The three lines share the same upper state so that, if the emitting atoms
were in an optically thin environment, the relative fluxes would be
in the ratio of the Einstein $A_{ij}$-values: $F_{1302}:F_{1304}:F_{1306}$ =
1.0 : 0.6 : 0.2. This expectation is not confirmed. Inspection of the spectrum
(Figure 1) shows clearly that the 1306\AA\ line is much stronger than the
1302\AA\ line, not markedly weaker as expected: the flux ratio is 1.7$\pm$0.2
not 0.2 (see also Table 1).
The 1304\AA\ blend appears 1.9$\pm$0.2 times stronger than the 1302\AA\ line.
If the Si II contribution to the blend is simply estimated from the
flux of the Si II line at 1309\AA\ and the  assumption that both lines are
optically thin, we can adjust the flux of the O I 1304\AA\ line. The
flux ratio relative to the 1302\AA\ line is then 1.4$\pm$0.2,
not the expected 0.6. 

To reconcile expected and observed fluxes, it is necessary to relax the
assumption that the gas is optically thin.  Several possible combinations
of geometrical and physical conditions may be considered. Three will be
mentioned here. First, suppose the emitting volume is itself optically thin
to the O I lines but an optically thick absorbing layer lies between us and the
emitting volume. If the absorbing gas is not cold ($T \geq$ 1000 K),
the relative populations of the three levels of the oxygen atom's
$^3$P$_{(2,1,0)}$ ground term will be given by the statistical weights ($g$)
and the line absorption coefficients will be in the ratio of the $gf$-values
and also the $A_{ij}$-values, as  the three lines share the same upper state.
In this scheme, if the  optical depth in the 1302\AA\ line is sufficiently
large, the line ratio will be reversed with the 1306\AA\ line stronger than
the 1302\AA\ line. Second, if the emitting region is optically thick but in
LTE, the emitted line fluxes, which are approximately equal, can be modified
by an absorbing layer but of lower optical depth than in the previous example.
In each case the required column density of the absorbing layer  could be
reduced considerably if the layer were cold. Then the line absorption
coefficient for 1302\AA\ line from the $^3$P$_2$ ground level could greatly 
exceed that for the 1306\AA\ line from the uppermost ($^3$P$_0$) level.
Since we have already mentioned the presence of an absorption feature
on top of the emission profile, it is obvious that some flux ratio reversal
is caused by radiative transfer under conditions of varying optical
depth and local temperature. Finally, the third possibility is that
the observed emission line fluxes might be generated by non-LTE effects.
One recalls the standard problem of a 2-level atom in an infinite slab (Mihalas
1978). Since the source function declines toward the boundary
($S_{\nu} = \sqrt{\epsilon}B_{\nu}$ in standard notation) and rises into
the slab saturating at $S_{\nu} = B_{\nu}$, an emission line shows
a central reversal (perhaps, unresolved at the low resolution of our spectra)
and lines of lower central depth will have higher intensities.

\subsection{Four Other Emitters: Si II, C II, S II, and N I}
Eight emission lines out of 19 are contributed by the Si II ion.
Here, as is the case with O I, we have line pairs that share an upper
state for their transitions: 1190 with 1194\AA\ from 10.40 eV;
1193 with 1197\AA\ from 10.38 eV; 1260 with 1264\AA\
from 9.83 eV; and 1304 (blended with O I) with 1309\AA\
from 9.49 eV. The typical spacing of the pairs (ca. 4.5 \AA) reflects
the two lower levels of Si II at 0.000 and 0.036 eV.
For the three pairs free of non-Si II blending, the $A_{ij}$ ratios of the
redder line to the bluer line are 5.0, 0.5, and 1.4. The corresponding observed
flux ratios are 1.2$\pm$0.3, 1.3$\pm$0.3, and 6.8$\pm$1.7, which confirm
the O I result that these
transitions are not emitted by an optically thin medium. With a $A_{ij}$
ratio of 2.0, it seems plausible to predict that 1309\AA\ should be stronger
than 1304\AA\ by at least a factor of 2, so that the latter's share of the flux
in its blend with O I is $\lesssim$ 25\%.

Other line pairs involve:
(1) two C II lines of multiplet 1, 1334\AA\  and a doublet at 1335\AA\,
with a combined $A_{ij}$ ratio of 0.2 + 1.2 = 1.4, observed flux
ratio is 3.1$\pm$0.8; (2) three S II multiplet 1 lines of a common lower
level with $A_{ij}$-ratios of 0.9 and 0.75, observed flux ratios are
1.8$\pm$0.3 and 0.5$\pm$0.1. The latter ratio cannot be considered as
optically thin because the 1259\AA\ is blended with at least one
absorption feature, and hence its actual strength and ratio are expected to
be higher; (3) two N I lines of multiplet 1 with a $A_{ij}$-ratio of 1.0,
observed flux ratio is 0.9$\pm$0.2.
This last coincidence of predicted and observed
flux ratios from N I indicate that perhaps this is the only species with
optically thin transitions. However, it seems odd that we
see only two N I lines, whereas three are expected. The missing middle line
at 1200.2\AA\ shares an identical $A_{ij}$ value with its detected
neighbors. We suspect that the absorption component of the 1200.7\AA\ line
is responsible for this discrepancy by overlapping the neighboring emission
line.

\section{INTERVENING ABSORPTION FEATURES}
\subsection{Absorption Companions of Emission Lines}
Our spectrum shows central
absorption components between the blue and red emission peaks (Figure 2).
In Table 2 we list all absorption features according to identified species
(compare to Table 1), excitation energy, rest wavelength, derived
radial velocity, observed equivalent width, and FWHM of the Gaussian profiles
that fit the features.
All our emission lines are transitions to ground and very
low-excitation states ($E_{j} \leq$ 0.04 eV)
and thus their absorption reversal is readily produced in a (cold) intervening
gas along the line of sight. We already discussed in \S3.2 and \S3.3 the
implication that the observed emission has been processed through a path
of high optical depth, possibly modified by intervening colder medium.
As mentioned above, averaging results from eight absorption features
yields a radial velocity of $-20\pm$14 km~s$^{-1}$. The lines
are narrow, with an average FWHM of 0.50$\pm$0.15 \AA, obviously
they are not resolved in this observation.

From Figure 2 and Table 2 it is apparent that a relationship exists between
the strength of the absorption (as determined by both species abundance and
transition strength) and the apparent radial velocity of the red
emission peak. More specifically, two C II and five Si II absorptions are
associated with red emission peaks at +115$\pm$12 and +116$\pm$11 km~s$^{-1}$,
respectively, whereas three S II lines with much weaker absorption
display a bluer emission peak at +104$\pm$4 km~s$^{-1}$. Furthermore,
the three O I lines which decrease in their relative transition strength
from 1302 to 1306 \AA, also decrease in radial velocity from +112 to +78
km~s$^{-1}$. Because of this shifting of emission peaks
and the added complication that different species may originate at different
physical locations in the optical medium and thus acquire different radial
velocities, the ``real'' velocity cannot be found without a detailed analysis
of the radiative transfer problem. For now, we resort to using average radial
velocities of line samples while keeping track of the errors involved.

\subsection{``Emission-Free'' Absorption}
We have identified seven transitions that appear in absorption only
in the spectrum of $\epsilon$~Aur, i.e. do not accompany emission lines.
These include six wider features composed of clusters of C I absorption lines
and a single strong absorption line ($W_{\lambda}$ = 900 m\AA) that is
the unresolved blend of the Mg II
transitions at 1239.93 and 1240.40 \AA. From our measurements we derive a
radial velocity of +12 km~s$^{-1}$ for Mg II under the assumption that the
feature's rest wavelength is represented by the average $\lambda_0$ of the
two transitions. This result puts Mg II in a class by itself, not
sharing any of the emission component velocities, nor that of the central
absorption ``on top'' on the emission. Again, more accurate velocity results
should be pursued via higher-resolution work.

Ground state and very low-excitation ($\leq$ 0.005 eV) levels of interstellar
C I appear to form six
clusters of blended absorption lines between 1189 and 1329 \AA. Thus we report
total cluster equivalent widths as follows: $\lambda$1189, 650 m\AA; 1193,
140; 1261, 250; 1277, 650; 1280, 460; and 1329, 670. Two of the C I clusters
are irretrievably blended with stellar emission lines, i.e.,
one cluster is under Si II at 1193 \AA, while another is under Si II 1260 \AA.
In both cases the absorption
is still visible to the red and blue of the emission line, the latter being
of anomalously weak flux due to attenuation by the C I feature (see Table 1).
It is therefore understandable why these two clusters have the two lowest
equivalent width values among all clusters.
Although the 1189 \AA\ feature lies adjacent to the emission line of
Si II 1190 \AA, the emission appears to be unaffected.

While the H I emission is of terrestrial origin, the entire contribution to
the strong Ly$\alpha$ absorption line is probably from interstellar absorption,
since the line is saturated down to zero residual intensity in the continuum,
with a typically flat bottom expected under such circumstances
(see, e.g., Diplas and Savage 1994).
According to Figure 1 of Diplas and Savage (1994), interstellar H I profiles
have FWHM of $\sim$10 and
$\sim$15 \AA, for H I column densities of 10$^{20.5}$ and 10$^{21.1}$ cm$^{-2}$,
respectively, as observed for stars $\sim$400 and $\sim$800 pc
away and at a galactic latitude of $\geq$ 13\fdg1.
Since $\epsilon$~Aur's galactic latitude is only 1\fdg2 and it
is believed to be at least 600 pc away (its parallax is given by {\it HIPPARCOS}
as 1.60$\pm$1.16 mas, see Perryman 1997), it is not surprising to find a FWHM
of 21$\pm$1 \AA\ in our spectrum. The inferred column density of H I is
definitely $\geq 10^{21.1}$ cm$^{-2}$, according to this comparison with
the stars of Diplas
and Savage.

\section{CONTINUUM FLUX LEVELS}
\subsection{{\it IUE\/} Archive Spectra}
In this section we shall inspect the overall spectral distribution of
$\epsilon$~Aur in order to try and separate the contributions from the
primary supergiant and from any UV source that is not occulted during primary
eclipse.
As mentioned before, such contribution should exist below $\sim$1400 \AA,
because in that region {\it IUE\/} observations revealed a diminished eclipse
signature (Parthasarathy and Lambert 1983 and references therein).
Since the primary star of the $\epsilon$~Aur system is usually thought
to be an F0 Ia supergiant, we have searched the {\it IUE\/} archive (King 1997)
for similar specimens.
The best {\it IUE\/} data (taken through the large aperture) sport a resolution
appreciably lower than that of the GHRS spectrum, namely, for the SWP, FWHM
is between
4.6---5.4 vs. 0.66 \AA. Spectra taken in the higher resolution mode of the
{\it IUE\/} suffer from lack of an appreciable signal in our region of interest.

For proper comparison of archival {\it IUE\/} spectra with the GHRS exposure
one has to degrade the {\it HST\/} data, i.e. transform it
into a lower resolution spectrum. {\it IUE\/} spectra of $\epsilon$~Aur
come with varying continuum slopes and flux levels, since even the primary
supergiant happens to be a variable star. For the purpose of this comparison
we chose the {\it IUE\/} exposure SWP 29696 (epoch: 1986 November 17) which has
one of the highest S/N ratios among $\epsilon$~Aur spectra obtained in the
years following the conclusion of the last eclipse.
In Figure 3 we show the {\it IUE\/}
spectrum of $\epsilon$~Aur and its resolution-degraded GHRS exposure.
The individual emission lines are no longer detected except for the
very strong O I/Si II complex at 1305\AA. We have changed the
flux values for the {\it IUE\/} exposure by $\times$0.63 for proper alignment
with our GHRS spectrum in Figure 3. Note that the sharp rise in {\it IUE\/}
flux below 1200 \AA\ is spurious.
Since the two spectra match pretty well, we conclude that the emission
lines could also be present in {\it IUE\/} spectra of earlier epochs, not only
in the recent GHRS spectrum. Such a good agreement between the spectra also
serves as an indicator that no major spectral contribution is attributable
to the hemisphere of the primary which is facing the secondary and which was
visible around primary eclipse, but is now definitely hidden from view.

\subsection{$\alpha$~Carinae and $\phi$~Cassiopeiae}
From the {\it IUE\/} archive we extracted short- and long-wavelength
spectra of $\alpha$~Car (HD 45348) and $\phi$~Cas (HD 7927), both well-known
F0 supergiant comparison stars of $\epsilon$~Aur. However, caution
should be exercised with $\alpha$~Car since its color index is bluer by
about 0.05 mag than that of other F0 giants and supergiants
observed with {\it IUE\/}, while it is intrinsically a fainter supergiant
with a luminosity class II, rather than I.

In Figure 4a we show the short-wavelength comparison between the low-resolution
version of our GHRS spectrum and both $\alpha$~Car and $\phi$~Cas, which are
extracted from the {\it IUE\/} exposures SWP 03382 and SWP 20288, respectively.
We have corrected the flux of $\phi$~Cas for interstellar reddening:
$E(B-V)$ = 0.50,
a value averaged from Bersier (1996) and Wolff, Nordsieck, and Nook (1996).
Thanks to its nearness, no significant interstellar reddening affects
the flux of $\alpha$~Car. Flux levels have been re-normalized
by $\times$0.21 and $\times$4.6, respectively, in order to equalize
their mean flux levels (over the interval 1600---2000 \AA) with that of
$\epsilon$~Aur. 

The long-wavelength exposures LWP 09537 ($\epsilon$~Aur) and LWR 16215
($\phi$~Cas) match very well below ca. 2900 \AA, especially considering
the differing amounts of dereddening applied to the two supergiants.
(There is no {\it IUE\/} spectrum for $\alpha$~Car at this low resolution.)
Based on mean flux values over 1850---3350 \AA,
the comparison $\phi$~Cas has been normalized by $\times$3.2 for proper
alignment with $\epsilon$~Aur. As seen in Fig. 4b,
we achieve a good match with $\epsilon$~Aur over the 1600---2900 \AA\
interval, with both stellar continua following very similar undulations.
Between 2900---3350 \AA\ the two
spectra are more separated from each other, but by switching roles in
being ``on top'' their average fluxes are in better agreement. This situation
happens in a region of lowest detector sensitivity for the LWP.

Below $\sim$1600\AA\ the spectrum of $\phi$~Cas is very
noisy due to very low signal which must be dominated by scattered light.
At longer wavelengths any flux from the secondary object of $\epsilon$~Aur
is insignificant relative to the brightness of the primary supergiant.
Other workers have documented and analyzed
detections of an excess flux, see e.g., Hack and Selvelli (1979) and
Parthasarathy and Lambert (1983), the latter using {\it IUE\/} spectra taken
outside and inside
of eclipse in order to decouple the contributions from the primary, secondary,
and scattered light in the spectrograph. Although we cannot at this time
deconvolve the GHRS spectrum into two contributions (from the primary and
secondary bodies), we confirm that the procedure used by
Parthasarathy and Lambert (1983) was correct. They were able to derive that
about 75\% of the observed flux level at 1250\AA\, i.e., about
7.5$\times$10$^{-13}$ erg cm$^{-2}$ s$^{-1}$ \AA$^{-1}$, was due to the
secondary (or un-eclipsed) source, with 25\% of the flux attributable to
{\it IUE\/} scattered light. Indeed, our GHRS spectrum is outstanding thanks
to absence of any scattered light, as revealed by the zero flux bottom of
the interstellar Ly$\alpha$ absorption line, and it shows that the flux
level at 1250\AA\ is, indeed, at 7.2$\times$10$^{-13}$ erg cm$^{-2}$ s$^{-1}$
\AA$^{-1}$, after being degraded to {\it IUE\/} resolution. We submit that
this should be the observable continuum from
the secondary, or any other source that is not occulted by the secondary
and its disk during primary eclipse.

\section{BINARY RAMIFICATIONS}
Thirteen years have elapsed since the last mid-eclipse time of $\epsilon$~Aur
until our GHRS spectrum was secured. That puts the secondary body of the system
squarely on the other side of its orbit, i.e., behind the plane of the sky
passing
through the primary. However, although 13 years also happen to be very close to
one-half of the period of the binary, 1996 was not expected to be the year of
the secondary eclipse due to the appreciable
eccentricity of the system (Morris 1962; Wright 1970).

\subsection{Orbital Kinematics}
The key quantity
is the mass ratio of the two stars, $q = M_1/M_2$, which fixes the ratio of
the radial velocity amplitudes, i.e., $q = K_2/K_1$, where $K_1 = 15.0\pm0.6$
km~s$^{-1}$ (Wright 1970).
On 1996 February 16, the orbital phase was 0.6971 from the time of periastron
(T = 1977 April 02 = JD 2443236). The primary supergiant's radial velocity
at the time was $v_1 = -$10.3 km~s$^{-1}$.
For sample mass ratios of $q = M_1/M_2$ = 1.0 and 2.0, one predicts secondary
radial velocities of $v_2$ = +7.5
and +16.4 km~s$^{-1}$, after correcting by the $\gamma$-velocity.

If the emission features are associated with the disk around
the secondary, there is the possibility that the secondary's radial
velocity can be derived from the emission lines. We have described how some
of the emission lines are split by an absorption line that originates
either in the ISM or even in the secondary's disk or other local cool gas.
This mix of absorption and emission complicates the use of the emission
as a speedometer. In particular, the red peak is demonstrably not providing
the secondary's radial velocity.
The observed radial velocity of the red peaks (+108 km~s$^{-1}$) yields a
mass ratio $q = v_2/v_1$ = 12$\pm2$ but this with the well-determined
mass function of 3.25$\pm$0.38 $M_{\sun}$ (Webbink 1985, who used Wright 1970)
gives stellar masses of 6600 and 550 $M_{\sun}$, respectively. These
extraordinary masses show that the red peak is not a tracer of the secondary's
velocity. Even more firmly, we reject the blue peak as a tracer because
the expected radial velocity of the secondary cannot be negative!
A chromospheric origin around the supergiant for the emission lines, which
was one solution suggested by Parthasarathy and Lambert (1983), calls for
emission centered on the smaller negative radial velocity of the primary,
but only our absorption component has an observed velocity closest to that
predicted for the primary.

The third option is to calculate the average velocity of
the peaks and use the result as our kinematic tracer of the secondary.
For this we assume that the source of the emission must be distributed
symmetrically around the secondary companion. We first compute the average
velocity of the two peaks for each transition
and then derive the final global average of kinematic
tracers. Seven pairs of red and blue emission peaks are available from
Tables 1 and 2 after rejecting N I at 1199.55 \AA\ due to severe blending.
From these emission pairs the global average of $v_2$ is +13.4$\pm$4.9
km~s$^{-1}$. Finally, as mentioned before, our
radial velocity scale includes a 4 km~s$^{-1}$ redshift as demonstrated
by the Ly$\alpha$ Geo-coronal line. When we correct the velocities above
for this shift, the tracer velocity is +9.4$\pm$4.9 km~s$^{-1}$, or $v_2$ =
+10.8$\pm$4.9 km~s$^{-1}$ after removal of the $\gamma$-velocity. This yields
a mass ratio of $q = 1.2\pm0.6$, i.e., two binary components of almost
equal mass, but with the primary as the slightly heavier object.
Indeed, component masses are 19$\pm$14 and 16$\pm$9 $M_{\sun}$ for the
primary and secondary, respectively. These masses correspond to the absorption
case with $p$ = $v_K/v_{rot}$ = 1.5 as prescribed by Lambert and
Sawyer (1986), although they preferred eventually to adopt $p \leq$ 1.2
and $q \le$ 0.4,
leading to mass limits of $M_1 \leq$ 3 and $M_2 \leq$ 6 $M_{\sun}$, i.e,
values that are just below the one $\sigma$ level of our results.

The masses just derived are reminiscent of the ``high-mass'' models in the
literature, pertaining to a massive F0 Ia supergiant and a very under-luminous
secondary. Our near-infrared spectroscopy shows that the pulsating primary
of $\epsilon$~Aur is spectroscopically indistinguishable from other massive
yellow supergiants such as $\phi$~Cas. But the secondary poses an enigma.
Lissauer and Backman (1984) have suggested the presence of a binary star
at the center of the disk, thus reducing the luminosity requirements. First,
a reduction by a factor of $\sim$10 is due to the smaller mass of the (new)
components; then, another reduction by a factor of $\sim$10 is suggested
by radiation escaping through the poles of the disk. Carroll et al. (1991)
prefer the high-mass scenario (using $M_2 \sim 14 M_{\sun}$) for their model
of a proto-planetary disk around $\epsilon$~Aur. More recently, Lissauer et al.
(1996) modeled the disk hydrostatically and concluded that its height
to radius ratio of $\lesssim$ 0.03 favors the presence of a massive secondary
with $M_2 \sim 15 M_{\sun}$.

\subsection{Discussing the Disk}
From the separation of blue and red emission peaks we derive a rotational
velocity of 103$\pm$20 km~s$^{-1}$ under the assumption of a circum-secondary
orbital motion. A Keplerian orbit for the gas implies a radius vector
of 1.4$\pm$1.0 AU away from a central point of 16$\pm$9 $M_{\sun}$.
This result is consistent with previous size estimates of the inner
transparent region of the disk. Carroll et al. (1991) model the hole
in the disk and find that its radius is 1.65 $R_{\rm primary}$ = 1.54 AU.
Furthermore, for a semi-major axis of $a$ = 13.35 AU (Webbink 1985)
we derive an annulus radius of 0.11($\pm0.07)a$, again a value very consistent
with the model by Wilson (1971) who found an inner disk radius of 0.1$a$.
Obviously the supported picture is of an opaque disk with an inner edge
irradiated and ionized by hot central source(s), thus producing emission
lines, while the cooler regions toward the outer edge are detectable via
absorption lines during primary eclipse.

One recalls the observations performed
during primary eclipse of absorption lines that changed from red shift to
blue shift by $\sim$66--84 km~s$^{-1}$ in a manner consistent with a rotating
disk (Lambert and Sawyer 1986, Sait\={o} et al. 1987).
Here we find from emission peak separation a radial velocity difference of
206 km~s$^{-1}$, which under Keplerian rotation shows that the radius vector
of the emitting region is approximately 6 to 10 times smaller than the
annulus responsible for the absorption lines. (Lambert and Sawyer gave
$v_{\rm rot}$ = 33 km~s$^{-1}$, Sait\={o} et al. gave 42 km~s$^{-1}$.)
This purely spectroscopic deduction is consistent with photometric models.
For example, the outer edge of the disk has been modeled to be 6.1 times greater
than its inner edge by Carrol et al. (1991). This modeled ratio
is derived from primary eclipse photometry in general and from the mid-eclipse
brightening in particular.

Another observational result is the unequal brightness of the two peaks,
a ratio of 3.7:1 in favor of the average red peak.
Theoretical disks with rotational symmetry should produce two peaks of equal
strength. In our case we already know about the central absorption, and
perhaps there is stronger absorption overlapping the blue peak
but which cannot be resolved in our spectrum. Another possibility may
involve a hot spot on the redshifted side of the disk, i.e., a stream
of gas that comes from the primary supergiant impacts onto the disk side
that is rotating away from us and the increased amount of matter and
gravitational energy result in a more pronounced red emission peak.
Unfortunately, the difficulty with this scenario would be the formation
of a hot spot by the {\it outer} edge of the disk, where the Keplerian
rotational velocity is much lower than our derived velocity.

\subsection{Eclipse Geometry}
As remarked earlier, the eclipse of the secondary object is due very soon,
unless it has started already!
From the orbital solution we find that the 1$^{st}$ and 4$^{th}$ contacts
will occur on ca. 1999 June 29 and 2000 December 11, respectively,
pending orbit-to-orbit variations in both eclipse
length and its mid-time. During secondary eclipse the optical brightness
of $\epsilon$~Aur will not change, being solely determined by the primary star.
Only in the UV can observers hope to detect
the secondary eclipse, perhaps in a most dramatic way when the emission lines
we detected should weaken in the spectrum of the system during ingress
and then, perhaps, vanish altogether during total secondary eclipse. Such a
remarkable transformation would greatly benefit the confirmation and/or
update of orbital elements and the determination of shapes and sizes for the
components in this unique binary star system. It is interesting to note that
had the orbital eccentricity been merely 0.0, there would have been a secondary
eclipse in progress already at the end of 1996 January, i.e., ca. three weeks
prior to the acquisition date of our GHRS spectrum. Furthermore, if the F0
supergiant happened to support a chromosphere with appreciable UV-continuum
attenuation, secondary eclipse would start earlier and last longer than
contact times calculated according to the photospheric size of the primary.
Incredibly, then, the relative weakness of the blue side of the emission
profiles can be explained also in terms of some eclipsing effect by the
primary star. The latter's
atmosphere could also be invoked as one of the possible sources for
the central absorption, because the two differ by about 10 km~s$^{-1}$, which
is smaller than the observational error. Such a possibility is exciting
for the option it offers to probe the outer atmosphere of the F0 supergiant.
Of course, as the secondary moves closer to the line of sight of the
primary, this effect should become stronger and easier to observe.

Secondary eclipse phases should be followed closely at higher resolution
with the STIS on board the {\it HST\/}, to provide better separation of
emission and absorption components and to measure the systemic velocity of
the gas more accurately. Later on,
following secondary eclipse, the opportunity to inspect the secondary on the
far side of the orbit would still be viable for a while, until $\epsilon$~Aur
starts heading towards its next primary eclipse, centered on about
2010 August 05. We note that using {\it HST\/} is the only way available
for obtaining the radial velocity curve of the secondary
and hence for finally determining the long-sought masses in this system.

\section{CONCLUSIONS}
In this paper we reported the detection of 17 optically-thick emission lines
in the {\it HST\/} spectrum of $\epsilon$~Aur, which belong to low-excitation
transitions in common atomic species. With a mean radial velocity of $\simeq$
+108 km~s$^{-1}$, these lines happen to be red peaks as part of wider emission
profiles. Weaker blue peaks give a mean radial velocity of $\simeq -$92
km~s$^{-1}$. The flux depression between the two peaks has a contribution
from an absorption component at $\simeq -$20 km~s$^{-1}$.

We interpret the double-peaked emission lines as rotational signatures of
the circum-secondary disk. From the average velocity of seven pairs of emission
peaks we derived a secondary radial velocity of $\approx$ +10.8 km s$^{-1}$,
a value which corresponds to a binary mass ratio of $q \approx$ 1.2, and
stellar masses of $M_1 \approx$ 19 and $M_2 \approx$ 16 $M_{\sun}$.
(All our derived values are accompanied by appreciable
error bars, as explicitly described above.) These are consistent with the
``high mass'' models of $\epsilon$~Aur---see, for example, Lissauer et al.
(1996) who model a disk surrounding a 15 $M_{\sun}$ secondary.
``Low-mass'' models (Sait\={o} et al. 1987, Lambert and Sawyer 1986) are
formally excluded by our results but the quoted uncertainties do not include
two likely key sources of additional uncertainty: our untested assumptions
that (i) the emission is centered on the secondary's orbital velocity, and (ii)
the appropriate velocity is the mean of red and blue emission peak velocities.
Therefore, we caution that the indication that ``high-mass'' models are
preferred should not be accepted as ending a long debate over $\epsilon$~Aur's
mass and evolutionary status. We offer the results as an indication that
the secondary mass is possibly obtainable from these emission lines if
they are observed at the higher resolution of STIS and observed sufficiently
often that an adequate segment of the secondary's velocity curve
can be defined.

From the velocity separation between seven pairs of emission peaks we derived
an average disk rotational velocity of $\simeq$ 103 km~s$^{-1}$, which
corresponds to an inner disk radius of $\approx$ 1.4 AU, or about six times
smaller than the outer regions of the disk as detected via absorption lines.
Not surprisingly, the emission is generated in a more energetic medium much
closer to the high-temperature secondary, at a disk radius which is also
about 10 times smaller than the primary's orbital semi-major axis.
Both the orbital and the rotational kinematic methods agree with each other
and point to a moderately massive binary system for $\epsilon$~Aur.

\acknowledgments
We have used the SIMBAD database during the writing of this paper. We
thank NASA for grant NAG 5-1616. We thank J. Tomkin for his
help with binary orbit analysis. YS thanks DLL and L. Trafton for providing
a life-saving post-doc employment. We thank an anonymous referee for a
helpful report.

\newpage
\figcaption[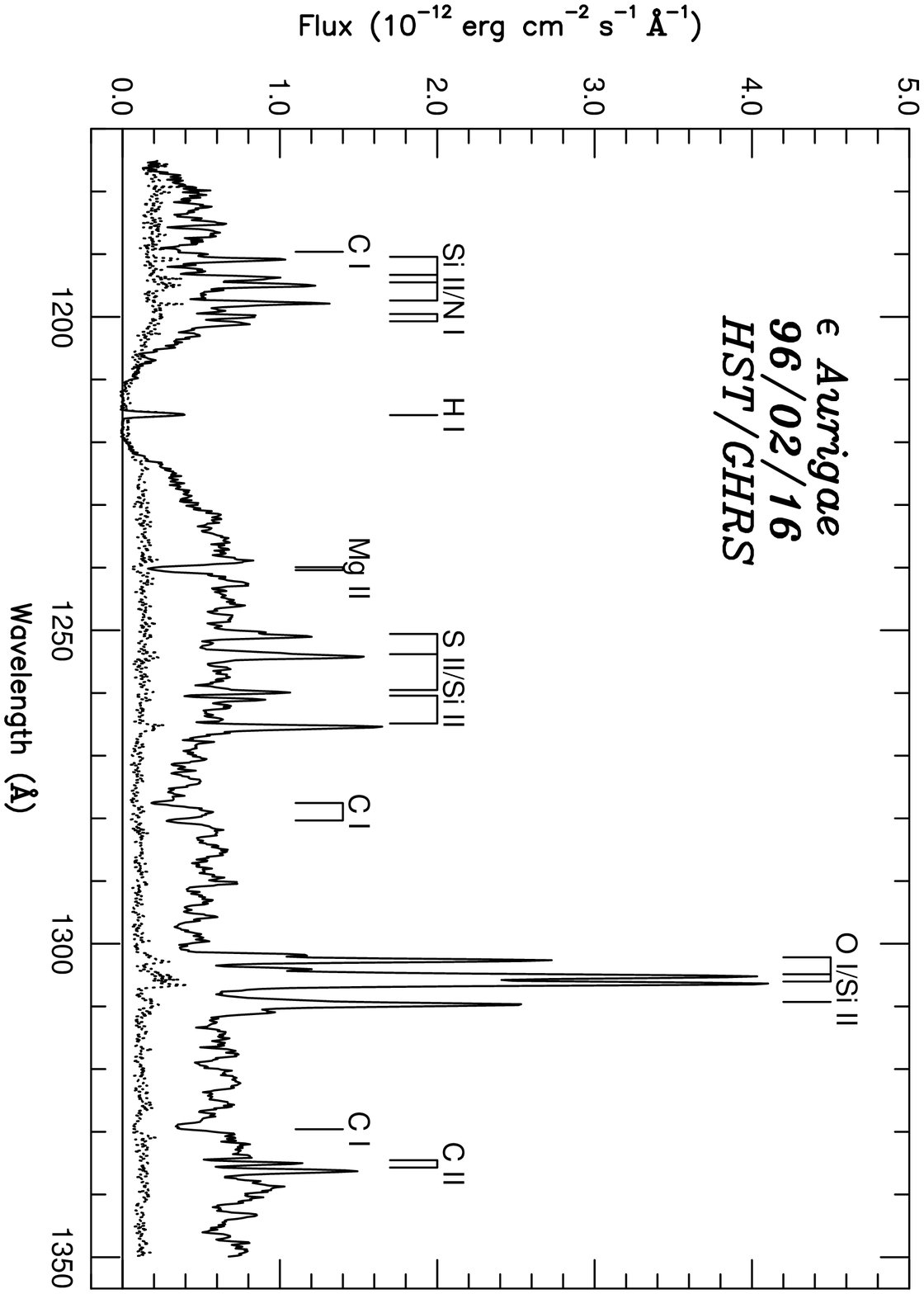]{The fully-reduced GHRS spectrum of $\epsilon$~Aur, after
correction for a reddening of $E(B-V)$ = 0.3. Emission lines and strong
absorption features are labeled in a higher and lower labels, respectively.
We do not show the reddest part of the spectrum (1350---1461 \AA) where
the photospheric absorption spectrum of the primary is dominant.
The lower dotted curve depicts the error
in the observed flux level, as derived by the calibration at STScI. Note
the Ly$\alpha$ emission line of H I which serves as as excellent standard for
profile shape and position for this spectrum. \label{fig1}}

\figcaption[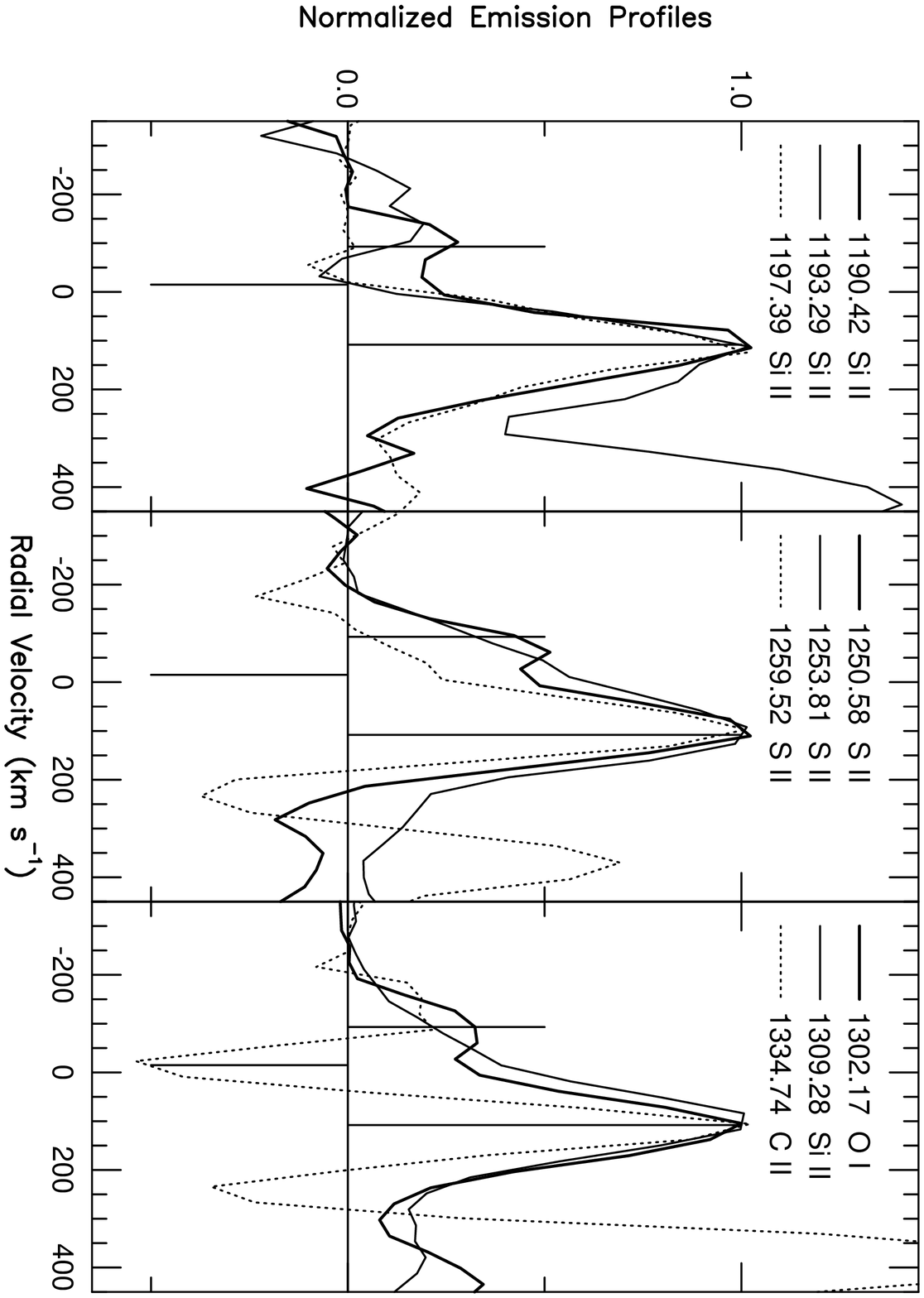]{A closer view of nine emission line profiles in
$\epsilon$~Aur
reveals their construction as a stronger red peak with a weaker blue bump and
a middle absorption feature. The average radial velocities of these three
components are marked by vertical lines.
All emission lines have been normalized by a continuum
subtraction (base at 0.0) and a division by their peak value (top at 1.0).
The left and center panels show common-multiplet transitions.
In the right panel, C II exhibits a very strong absorption feature
and a red peak that is more redshifted than the average radial velocity of
+108 km~s$^{-1}$. Note that the red wings of the lines at 1193, 1259, and
1334 \AA\ are affected by some blending with their redder neighbors, i.e.,
$\lambda\lambda$ 1194, 1260, and 1335. \label{fig2}}

\figcaption[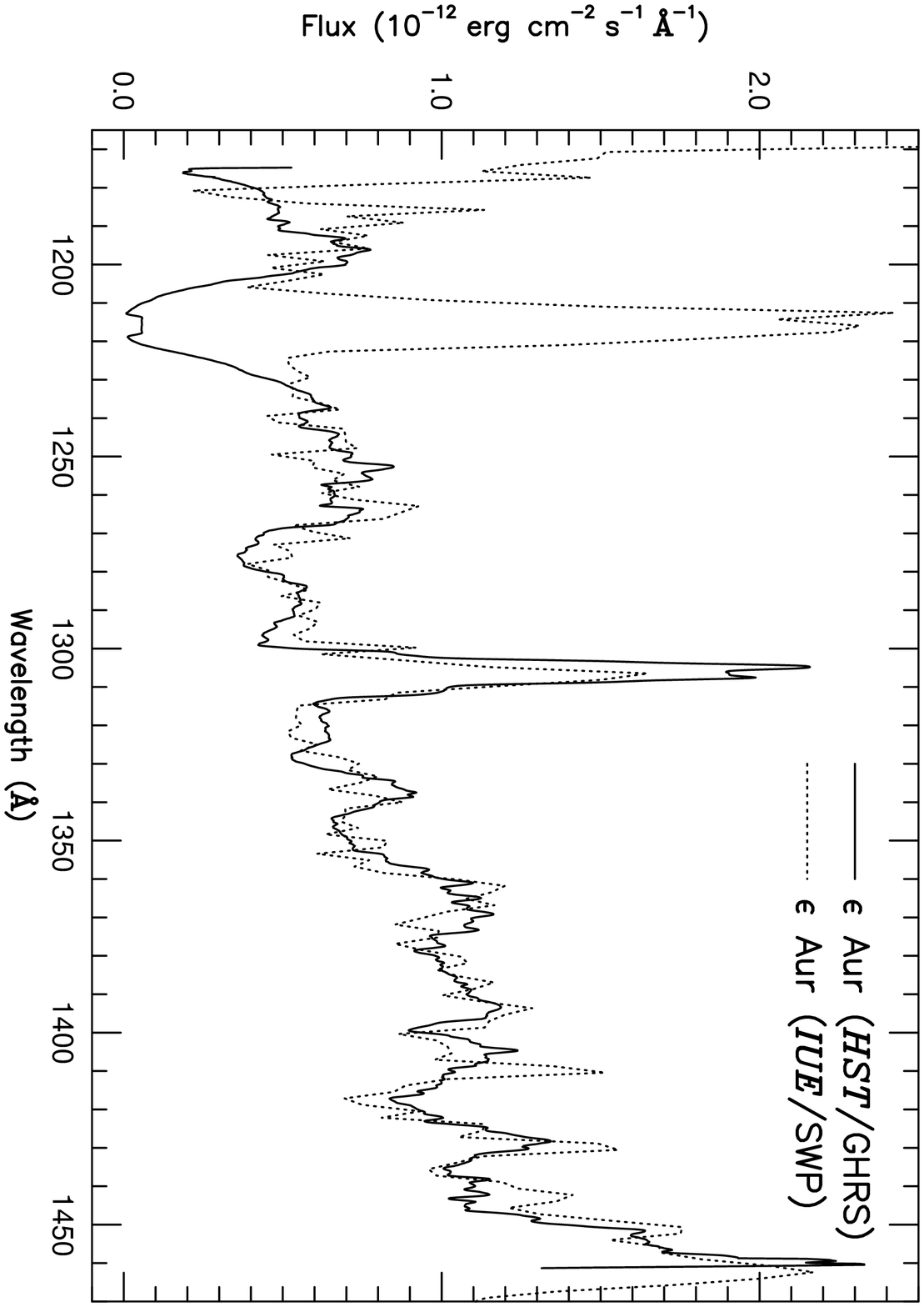]{A comparison of our GHRS spectrum, degraded to lower
resolution of 5 \AA\ by a 35-pixel box smoothing, and the {\it IUE\/}
exposure SWP 29696 from ten years earlier.
The latter has been re-scaled by $\times$0.63
for proper alignment. Both spectra have been corrected for interstellar
reddening. \label{fig3}}

\figcaption[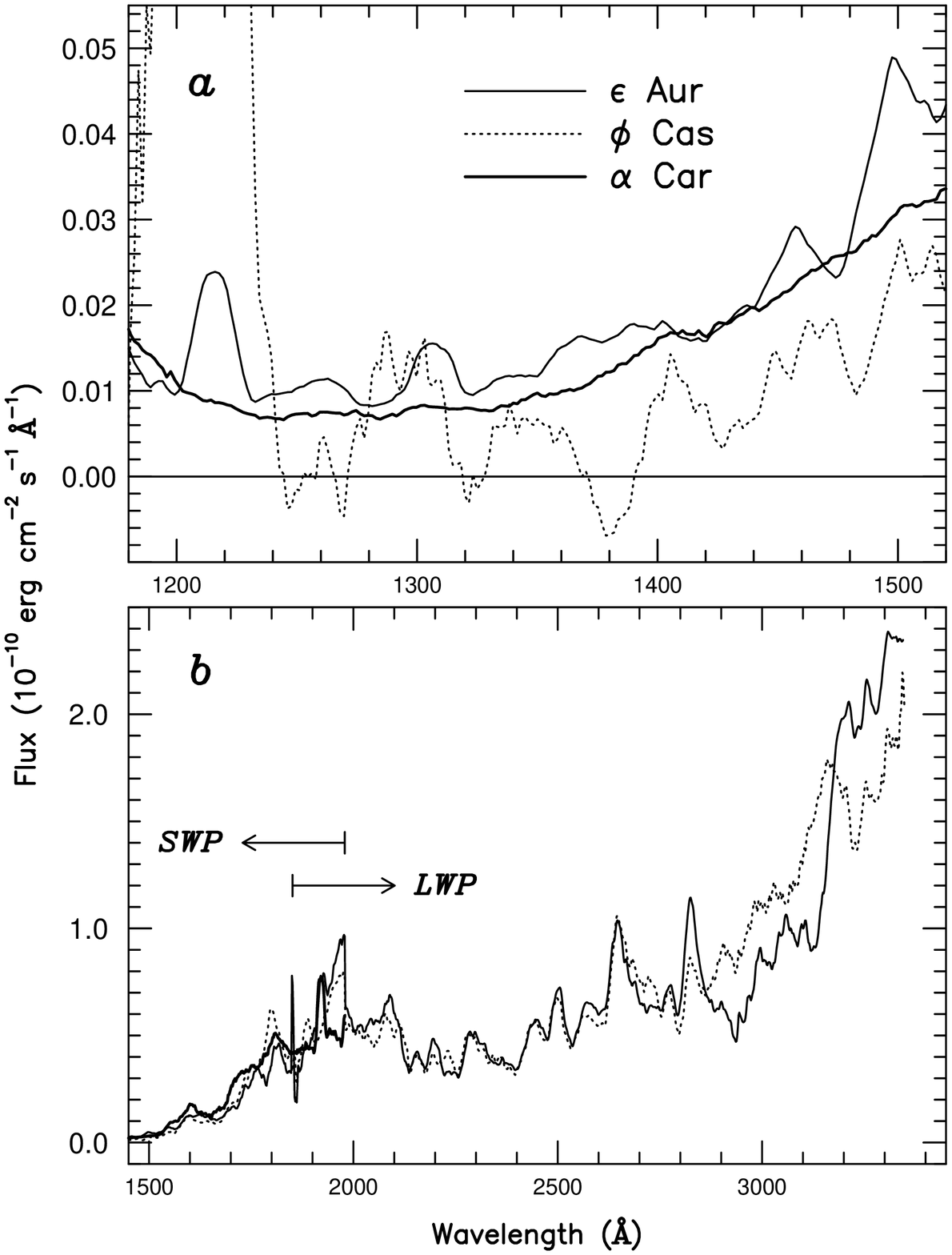]{Three F0 supergiants observed by the {\it IUE\/} are
compared here after combining their SWP and LWP spectra.
Panel {\it a} shows the
shorter wavelength region that corresponds to the coverage of our GHRS
exposure. In panel {\it b} the longer wavelength region is shown. The match
is very good below
2900 \AA. A small region around 1900 \AA\ suffers from a mis-match
between the SWP and LWP independent exposures. \label{fig4}}

\begin{deluxetable}{lcccrrrr}
\small
\tablewidth{0pt}
\tablecaption{Red Emission Peaks of $\epsilon$~Aur \label{tab1}}
\tablehead{
\colhead{Atom}
&\colhead{Multiplet}
&\colhead{$\lambda_0$}
&\colhead{$\lambda_{\rm obs}$}
&\colhead{$v_{\rm rad}$}
&\colhead{$A_{ij}$}
&\colhead{Flux\tablenotemark{a}}
&\colhead{FWHM}
\\
\colhead{}
&\colhead{}
&\colhead{(\AA)}
&\colhead{(\AA)}
&\colhead{(km~s$^{-1}$)}
&\colhead{ns$^{-1}$}
&\colhead{}
&\colhead{(\AA)}
}
\startdata
Si II & (5) & 1190.42 & 1190.83 & +114 & 1.539 & 3.8$\pm0.7$ & 0.58 \nl
Si II & (5) & 1193.29 & 1193.81 & +131 & 6.103 & 3.4$\pm0.4$ & 0.80 \nl
Si II & (5) & 1194.50 & 1194.98 & +121 & 7.652 & 4.7$\pm1.0$ & 0.71 \nl
Si II & (5) & 1197.39 & 1197.85 & +115 & 3.038 & 4.3$\pm0.7$ & 0.68 \nl
N I   & (1) & 1199.55 & 1199.93 &  +86 & 0.399 & 1.6$\pm0.2$ & 0.57 \nl
N I   & (1) & 1200.71 & 1201.09 &  +92 & 0.398 & 1.5$\pm0.2$ & 0.57 \nl
H I   & (1) & 1215.67 & 1215.58 &$-$22 & 0.939 & 2.8$\pm2.5$ & 0.65 \nl
S II  & (1) & 1250.58 & 1251.00 & +101 & 0.047 & 4.1$\pm0.5$ & 0.58 \nl
S II  & (1) & 1253.81 & 1254.26 & +108 & 0.042 & 7.2$\pm0.7$ & 0.55 \nl
S II  & (1) & 1259.52 & 1259.95 & +102 & 0.034 & 2.1$\pm0.4$ & 0.51 \nl
Si II & (4) & 1260.42 & 1261.07 & +155 & 2.066 & 1.0$\pm0.2$ & 0.47 \nl
Si II & (4) & 1264.74 & 1265.37 & +150 & 2.489 & 6.8$\pm1.0$ & 0.93 \nl
O I   & (2) & 1302.17 & 1302.66 & +112 & 0.315 & 19.2$\pm1.9$ & 0.72 \nl
Si II & (3) & 1304.37 & \nodata &\nodata&0.741 &\nodata&\nodata \nl
O I   & (2) & 1304.86 & 1305.23 &  +90 & 0.188 & 35.8$\pm2.4$ & 0.82 \nl
O I   & (2) & 1306.03 & 1306.38 &  +78 & 0.063 & 32.3$\pm2.1$ & 0.77 \nl
Si II & (3) & 1309.28 & 1309.72 & +101 & 1.467 & 19.3$\pm1.2$ & 0.71 \nl
C II  & (1) & 1334.53 & 1335.00 & +106 & 0.225 & 1.2$\pm0.3$ & 0.63 \nl
C II  & (1) & 1335.71 & 1336.26 & +123 & 0.273 & 3.7$\pm0.4$ & 0.65 \nl
\enddata
\tablenotetext{a}{in units of 10$^{-13}$ erg cm$^{-2}$ s$^{-1}$}
\end{deluxetable}

\begin{deluxetable}{lccrccrrc}
\small
\tablewidth{0pt}
\tablecaption{Blue Emission Peaks and Central Absorption of
$\epsilon$~Aur \label{tab2}}
\tablehead{
\colhead{Atom}
&\colhead{$E_{j}$}
&\colhead{$\lambda_0$}
&\colhead{$v_{\rm rad}^{\rm Blue}$}
&\colhead{\case{F(Blue)}{F(Red)}}
&\colhead{FWHM}
&\colhead{$v_{\rm rad}^{\rm Abs}$}
&\colhead{$W_\lambda$}
&\colhead{FWHM}
\\
\colhead{}
&\colhead{(eV)}
&\colhead{(\AA)}
&\colhead{(km~s$^{-1}$)}
&\colhead{}
&\colhead{(\AA)}
&\colhead{(km~s$^{-1}$)}
&\colhead{(m\AA)}
&\colhead{(\AA)}
}
\startdata
Si II& 0.00&1190.42&$-85$  &0.26   &0.58   &$-11$&100&0.58\nl
Si II& 0.00&1193.29&$-121$ &0.14   &0.41   &$-33$&180&0.44\nl
Si II& 0.04&1197.39&\nodata&\nodata&\nodata&$-40$& 69&0.19\nl
N I  & 0.00&1199.55&$-110$ &0.34   &0.57   &$-27$&153&0.57\nl
S II & 0.00&1250.58&$-62$  &0.52   &0.57   &$-28$& 34&0.43\nl
Si II& 0.04&1264.74&$-118$ &0.07   &0.63   &\nodata&\nodata&\nodata\nl
O I  & 0.00&1302.17&$-84$  &0.28   &0.56   &$-3$ &320&0.58\nl
Si II& 0.00&1304.37&$-87$  &\nodata&0.47   &\nodata&\nodata&\nodata\nl
Si II& 0.04&1309.28&$-69$  &0.21   &0.71   &\nodata&\nodata&\nodata\nl
C II & 0.00&1334.53&$-89$  &0.31   &0.54   & 0   &320&0.61\nl
C II & 0.01&1335.71&\nodata&\nodata&\nodata&$-16$&210&0.62\nl
\enddata
\end{deluxetable}
\end{document}